\documentclass[12pt]{article}
\usepackage{epsfig}
\usepackage{amsfonts}
\usepackage{amsopn}
\usepackage{amsmath}

\title{M-brane dynamical symmetry and quantization}
\author{
Jens Hoppe \thanks{e-mail: hoppe@math.kth.se}  \\
Department of Mathematics,\\
Royal Institute of Technology, \\
KTH, 100 44 Stockholm, \\
Sweden
}

\begin{document}
\date{}
\maketitle

\abstract{The recently discovered dynamical symmetry for
relativistic extended objects is derived from first principles, and
analogous commutators are obtained for the corresponding formal quantum expressions}

\section*{}
As shown in [1],  Lorentz invariance implies the existence of a
dynamical symmetry for M-branes, irrespective of the dimension; see
also [4,6]. Deriving the corresponding commutation relations
\emph{directly} (rather then deducing them from a simple, though indirect, argument) is tedious, but turns out to reveal very
interesting  commutation relations for the various objects involved:
the transverse embedding coordinates $x_i$ and their canonical
momenta $p_i$ ( constrained by $\int f^a \vec{p}\partial_a \vec{x}$
being zero whenever $\nabla_a f^a=0$), $x_{-}$ (here called $\zeta$; in
the light-cone formulation to be reconstructed from $\vec{x}$ and
$\vec{p}$) and the Hamiltonian density $\tilde{\mathcal{H}}$. 
The main object of interest is \footnote{for background
details, see [1-6]; also note that I will only at the
beginning write out the (unit weight)
non-dynamical density $\rho$ with respect to which (non-constant)
orthonormal (eigen) functions (of a Laplacian $\Delta$ on the
parameter space), $\{Y_{\alpha}\}_{\alpha=1}^{\infty}$, are
introduced}

\begin{equation}
M_{i-}:=\int (x_i\tilde{ \mathcal{H}}-\tilde{\zeta}p_i)d^M\varphi  \label{1}
\end{equation}
\[
x_i=X_i+x_{i \alpha}Y_{\alpha}, \ \ \ \ p_i=\rho P_i + \tilde{p}_i
\]
\[
\tilde{\zeta}= Y_{\alpha}(d_{\alpha \beta \gamma}+e_{\alpha \beta \gamma})\vec{x}_{\beta}\vec{p}_{\gamma}=2 \sum_{\alpha=1}^{\infty} \frac{Y_{\alpha}}{\mu_{\alpha}}\int (\nabla^a Y_{\alpha})\tilde{\vec{p}}\partial_a \vec{x} d^M\varphi
\]
\[
d_{\alpha \beta \gamma}=\int Y_{\alpha}Y_{\beta}Y_{\gamma} \rho d^M \varphi, \ \ \ e_{\alpha \beta \gamma}=\frac{\mu_{\beta}-\mu_{\gamma}}{\mu_{\alpha}}d_{\alpha \beta \gamma}
\]
\[
\frac{\tilde{\mathcal{H}}}{\rho}=Y_{\alpha} d_{\alpha \beta \gamma} \vec{p}_{\beta}\cdot \vec{p}_{\gamma}+Y_{\alpha}g_{\alpha}, \ \ \ \ \Delta Y_{\alpha}=-\mu_{\alpha} Y_{\alpha},
\]
\[
\frac{g}{\rho^2}=\frac{1}{\rho^2}det(\partial_a \vec{x} \partial_b \vec{x})_{a,b=1,\ldots,M}=g_{\alpha}Y_{\alpha}+\int g/\rho.
\]
One of the crucial relations will be
\begin{equation}  \label{2}
\{\tilde{\zeta}_{\alpha},\tilde{\mathcal{H}}_{\beta}\}= (3d_{\alpha \beta \gamma}+e_{\alpha \beta \gamma})\tilde{\mathcal{H}}_{\gamma}+ 4 \delta_{\alpha \beta} \mathbb{M}^2
\end{equation}
where
\begin{equation}  \label{3}
\mathbb{M}^2= \vec{p}_{\alpha} \vec{p}_{\alpha}+ \int g/\rho d^M \varphi
\end{equation}
denotes the squared mass (a relativistically invariant quantity) of
the extended object.

\section{Poisson Brackets}

When obtaining (\ref{2}) (and the other
relations stated below), I found it easiest to artificially
(re)introduce zero-modes $X_i$, $P_j$, $\eta$, $\zeta_0$, writing
\[
\tilde{\zeta}=2\eta (\zeta-\zeta_0)-2\vec{P}(\vec{x}-\vec{X})
\]
\begin{equation}
\tilde{\mathcal{H}}=2\eta(\mathcal{H}-H)-2\vec{P}(\vec{p}-\vec{P})  \label{4}
\end{equation}
with
\begin{equation}  \label{5}
\{\eta,\zeta_0\}=1, \ \ \ \{x_i(\varphi),p_j(\tilde{\varphi})\}=\delta_{ij}\delta(\varphi,\tilde{\varphi})
\end{equation}

\begin{equation}  \label{6}
\partial_a \zeta = \frac{\vec{p}\partial_a \vec{x}}{\eta \rho}, \ \ \ \ \dot{\zeta}= \frac{\vec{p}^2+g}{2\eta^2 \rho^2}.
\end{equation}
So (on the r.h.s. always working modulo zero-modes, as the l.h.s.
does not  contain any; hence knowing that they \emph{have} to
cancel/drop out) e.g. (from now on not writing out factors of
$\rho$)
\[
\{\tilde{\zeta}_{\alpha}, \tilde{T}_{\epsilon}=d_{\epsilon \beta \gamma}\vec{p}_{\beta}\vec{p}_{\gamma}\}= 2\{\eta \zeta_{\alpha}-\vec{P}\vec{x}_{\alpha},\int Y_{\epsilon}{\vec{p}}^{ \ 2}-2\vec{P}\vec{p}_{\epsilon}\}
\]
\[
\longrightarrow \frac{2}{\mu_{\alpha}}\{ \int (\nabla^a Y_{\alpha})\vec{p}\partial_a \vec{x}, \int Y_{\epsilon}\vec{p}^2 \}=-\frac{4}{\mu_{\alpha}}\int \nabla_a((\nabla^a Y_{\alpha})\vec{p})Y_{\epsilon}\vec{p}
\]
\[
=-\frac{4}{\mu_{\alpha}}\int \left(-\mu_{\alpha} Y_{\alpha}\vec{p}+\nabla^a Y_{\alpha}\nabla_a \vec{p} \right)Y_{\epsilon}\vec{p}=4\int Y_{\alpha}Y_{\epsilon} \vec{p}^2-\frac{2}{\mu_{\alpha}}\int (\nabla^a Y_{\alpha})Y_{\epsilon}\partial_a Y_{\beta}\int Y_{\beta}\vec{p}^2
\]
\[
\Longrightarrow 4 \delta_{\alpha \epsilon} \vec{p}_{\gamma} \vec{p}_{\gamma} + (e_{\alpha \epsilon \beta }-d_{\alpha \epsilon \beta})\tilde{T}_{\beta}+ 4 d_{\alpha\epsilon \beta} \tilde{T}_{\beta}
\]
\begin{equation}  \label{7}
= (3d_{\alpha \epsilon \beta}+e_{\alpha \epsilon \beta})\tilde{T}_{\beta}+4\delta_{\alpha \epsilon}\vec{p}_{\gamma} \vec{p}_{\gamma} = \{\tilde{\zeta}_{\alpha},\tilde{T}_{\epsilon}\}
\end{equation}
Similarly,
\[
\{\tilde{\zeta}_{\alpha},g_{\beta}\}=\{\frac{2}{\mu_{\alpha}}\int (\nabla^a Y_{\alpha})\tilde{\vec{p}}\partial_a \vec{x},\int Y_{\beta}g\}=\frac{4}{\mu_{\alpha}}\int(\nabla^c Y_{\alpha})\partial_c \vec{x} \partial_a (Y_{\beta}g g^{ab}\partial_b \vec{x})
\]
\[
=4\int Y_{\alpha}Y_{\beta}g - \frac{4}{\mu_{\alpha}}\int \nabla^c Y_{\alpha}Y_{\beta}g g^{ab}\underbrace{\partial^2_{ac}\vec{x}\partial_b \vec{x}}_{=\frac{1}{2}\partial_c g_{ab}}
\]
\[
=4\delta_{\alpha\beta}\int g +4d_{\alpha\beta\gamma}g_{\gamma}+\frac{2}{\mu_{\alpha}}\int \nabla_c (\nabla^cY_{\alpha}Y_{\beta})g
\]
\[
=4\delta_{\alpha\beta}\int g + 4d_{\alpha\beta\gamma}g_{\gamma}-2d_{\alpha\beta\epsilon}g_{\epsilon}+
\underbrace{\frac{2}{\mu_{\alpha}}\int(\nabla^cY_{\alpha})Y_{\epsilon}\nabla_cY_{\beta}g_{\epsilon}}_{=-(e_{\alpha\epsilon\beta}-d_{\alpha\epsilon\beta})g_{\epsilon}=(+e_{\alpha\beta\epsilon}+d_{\alpha\beta\epsilon})g_{\epsilon}}
\]
\[
=4\delta_{\alpha\beta}\int g+3d_{\alpha\beta\epsilon}g_{\epsilon}+e_{\alpha\beta\epsilon}g_{\epsilon}=\{\tilde{\zeta}_{\alpha},g_{\beta}\}   ,
\]  
proving (\ref{2}).

In the same manner one can see that
\begin{equation} \label{8}
\ x_{i\alpha}x_{j\alpha'}\{\tilde{\mathcal{H}}_{\alpha},\tilde{\mathcal{H}}_{\alpha'}\}+((\tilde{\zeta}_{\alpha})x_{i\beta}\{p_{j\alpha},\tilde{\mathcal{H}}_{\beta}\}-(i\leftrightarrow j))=0\
\end{equation}
and
\begin{equation} \label{9}
\{\int \tilde{\zeta}p_i,\int \tilde{\zeta}p_j\}=0.
\end{equation}
Namely,
\[
\{\tilde{\mathcal{H}}_{\alpha},\tilde{\mathcal{H}}_{\alpha'}\}=\{\int Y_{\alpha}(2\eta(\mathcal{H}-H)-2\vec{P}(\vec{p}-\vec{P})),\int Y_{\alpha'}(2\eta(\mathcal{H}-H)-2\vec{P}(\vec{p}-\vec{P}))\}
\]
\[
\rightarrow 4 \eta^2 \{\int\mathcal{H}_{\alpha},\int\mathcal{H}_{\alpha'}\}
\]
\begin{equation} \label{10}
\rightarrow 4 \int (Y_{\alpha}\partial_a Y_{\alpha'}-Y_{\alpha'}\partial_a Y_{\alpha})gg^{ab} \underbrace{\tilde{\vec{p}}\partial_b \vec{x}}_{=\frac{1}{2}\partial_b \tilde{\zeta}} ,
\end{equation}
hence
\begin{equation} \label{11}
\int \int x_i(\varphi) x_j(\tilde{\varphi}) \{\tilde{\mathcal{H}}(\varphi),\tilde{\mathcal{H}}(\tilde{\varphi})\} \approx 2\int (x_i \partial_a x_j - x_j \partial_a x_i)gg^{ab}\partial_b \tilde{\zeta} \rho d^M\varphi
\end{equation}
The remaining term(s) in (\ref{8}) however, are
\[
\int \tilde{\zeta}(\tilde{\varphi})x_i(\varphi)\{p_j(\tilde{\varphi}),g(\varphi)\}-(i \leftrightarrow j)
\]
\[
=+2\int \tilde{\zeta}(\varphi)\partial_a (x_igg^{ab}\partial_b x_j)-(i \leftrightarrow j)
\]
\begin{equation} \label{12}
=-2\int \partial_a \tilde{\zeta} x_i gg^{ab}\partial_b x_j -(i \leftrightarrow j),
\end{equation}
canceling (\ref{11}) (hence proving (\ref{8})).
 (\ref{9}) easily follows when using
\begin{equation} \label{13}
\{\tilde{\zeta}_{\alpha},\vec{p}_{\beta}\}=(d_{\alpha\beta\gamma}+e_{\alpha\beta\gamma})\vec{p}_{\gamma}
\end{equation}
and (already proven in [1], see also [3])
\begin{equation} \label{14}
\{\tilde{\zeta}_{\alpha},\tilde{\zeta}_{\beta}\} \approx 2 e_{[\alpha\beta]\gamma}\tilde{\zeta}_{\gamma}
\end{equation}
-the relations generalizing those of the Witt-algebra of the string
to arbitrary dimensions. The only missing part of having thus
(directly) proven
\begin{equation} \label{15}
\{M_{i-},M_{j-}\}=4 \mathbb{M}^2 M_{ij}, \ \ \ \ M_{ij}:=x_{i\alpha}p_{j\alpha}-x_{j\alpha}p_{i\beta}
\end{equation}
is the observation that
\begin{equation} \label{16}
\{\tilde{\zeta}_{\alpha},\vec{x}_{\beta}\}=(-d_{\alpha\beta\gamma}+e_{\alpha\beta\gamma})\vec{x}_{\gamma}
\end{equation}
implies
\[
p_{j\alpha}\tilde{\mathcal{H}}_{\beta}\{\tilde{\zeta}_{\alpha},x_{i\beta}\}- (i \leftrightarrow j)
\]
\begin{equation} \label{17}
=(x_{i\gamma}p_{j\alpha}-x_{j\gamma}p_{i\alpha})(-d_{\alpha\beta\gamma}+e_{\alpha\beta\gamma})\tilde{\mathcal{H}}_{\beta},
\end{equation}
and that
\begin{equation} \label{18}
x_{i\alpha}\tilde{\mathcal{H}}_{\alpha'}\{d_{\alpha\beta\gamma}\vec{p}_{\beta}\vec{p}_{\gamma},x_{j\alpha'}\}-(i \leftrightarrow j) = -2d_{\alpha\alpha'\gamma}(x_{i\alpha}p_{j\gamma}- x_{j\alpha}p_{i\gamma})\tilde{\mathcal{H}}_{\alpha'}
\end{equation}
i.e. (\ref{17})+(\ref{18}) canceling both the d-terms, as well as the e-term in $p_{j\alpha}x_{i\beta}\{\tilde{\zeta}_{\alpha},\tilde{\mathcal{H}}_{\beta}\}-(i \leftrightarrow j)$.

Coming back to (\ref{2}), note that (in particular) for fixed $\alpha=\beta$ one has
\begin{equation}  \label{19}
\{\tilde{\zeta}_{\alpha},\tilde{\mathcal{H}}_{\alpha}\}=4 \mathbb{M}^2 +  (3d_{\alpha \alpha \gamma}+e_{\alpha\alpha\gamma})\tilde{\mathcal{H}}_{\gamma}
\end{equation}
and that one also has (for arbitrary, fixed, $\alpha$)
\[
\{\tilde{\zeta}_{\alpha},\mathbb{M}^2\}=2\tilde{\mathcal{H}}_{\alpha}, \ \ \ \ \{\tilde{\zeta}_{\alpha},\{\tilde{\zeta}_{\alpha},\mathbb{M}^2\}\}=4\mathbb{M}^2+(3d_{\alpha \alpha \gamma}+e_{\alpha\alpha\gamma})\tilde{\mathcal{H}}_{\gamma}
\]
\[
\{\tilde{\vec{x}}_{\alpha},\mathbb{M}^2 \} = 2\vec{p}_{\alpha}, \ \ \ \  \{\tilde{\vec{p}}_{\alpha},\mathbb{M}^2 \} = 2(\Delta\vec{x})_{\alpha}
\]

\section{Commutators}

To derive corresponding relations for a quantized theory of
relativistic extended objects is very difficult; though \emph{formally} possible, when
organized with the help of the equations derived above (in
particular: (\ref{2}), (\ref{8}), (\ref{9}), (\ref{11})-(\ref{18})).
There are several differences compared to the classical case:
\begin{itemize}
\item an ordering prescription has to be chosen at the beginning for the generators to be (at least formally) hermitean
\item the expressions obtained after using the basic commutation relations quite often have to be reordered (in order to allow for cancelations present in the classical calculations)
\item special care has to taken when using the constraints
\end{itemize}
The final step, as essential as difficult (namely: renormalizing
/ making sure that the expressions one works with are well defined for
arbitrary M (in string theory furnished by normal ordering of the oscillator
modes; for M=2 one could e.g. combine the calculations below with finite N matrix
regularization)/ remains.

Before going into details, concerning
\begin{equation}  \label{20}
[\int(x_i \mathcal{H}+ \mathcal{H}x_i-\zeta p_i -p_i \zeta),\int(x_j \mathcal{H}+ \mathcal{H}x_j-\zeta p_j -p_j \zeta)]
\end{equation}
let me remark that there are, despite the high degree of non-linearity (from $\mathcal{H}$), and non-locality (from $\zeta$) a number of \emph{simplifying} features:
\begin{itemize}
\item $\mathcal{H}$ is a sum of "pure" terms (one depending only on the momenta, the other only on the coordinates)
\item as $\zeta$ is linear in $x$ and $p$, its naive commutator with any "pure" term (and also with itself) does not involve any complicated ordering questions.
All relevant Poisson-brackets involving $\zeta$ easily carry over to (formal) quantum relations (cp.[3])
\end{itemize}

As the definition of the purely internal quantum $\zeta$ let us take
\begin{equation}  \label{21}
\zeta_{\alpha}=\frac{1}{2}(d_{\alpha\beta\gamma}+e_{\alpha\beta\gamma})
(\vec{x}_{\beta}\vec{p}_{\gamma}+\vec{p}_{\gamma}\vec{x}_{\beta})+z_{\alpha}
\end{equation}
-which is formally hermitean (the real constants $z_{\alpha}$ are
put in for safety, resp. generality; due to the infinite sums
care must be taken concerning possible divergencies 

The classical expression [7]
\begin{equation}  \label{22}
\zeta=\zeta_0 + \frac{1}{\eta}\int G(\varphi,\tilde{\varphi})\tilde{\nabla}^a
(\frac{\vec{p}}{\rho}\partial_a \vec{x})
\end{equation}
implies ("$\rho=1$")
\[
\partial_c(\eta \zeta) = \vec{p}\partial_c \vec{x} - \int (\partial_c \tilde{\nabla}^aG +\delta^a_c \delta(\varphi,\tilde{\varphi}))\vec{p}\partial_a \vec{x}
\]
\begin{equation}  \label{23}
=\vec{p}\partial_c \vec{x} - \int F^a_c(\varphi,\tilde{\varphi})\vec{p}\tilde{\partial}_a\vec{x} d^M\tilde{\varphi} \approx \vec{p}\partial_c \vec{x}
\end{equation}
as $F$ is divergence-free,  $\tilde{\nabla}_a F^a_c(\varphi,\tilde{\varphi})=0$.
Similarly one may take $\hat{\zeta}:=\zeta_{\alpha}Y_{\alpha}(\varphi)$ to satisfy
\begin{equation}  \label{24}
\partial_c \hat{\zeta}=\left(\tilde\vec{p}\partial_c \vec{x} - \int F^a_c \vec{p}\partial_a \vec{x} \right) + h.c. + \partial_c z.
\end{equation}
Just as in eq. (25) of [1] (there for the classical case) it is easy to argue that
\[
-i[\zeta_{\alpha},\zeta_{\alpha'}]=\int Y^c_{[\alpha\alpha']}\tilde\vec{p}\partial_c \vec{x}
d^M \varphi + h.c.
\]
\[
=\int Y^c_{[\alpha\alpha']}(\partial_c\zeta +(\int F_c^a(\vec{p}\partial_a\vec{x})+h.c.)-\partial_c z )
\]
\begin{equation}  \label{25}
=2e_{[\alpha\alpha']\epsilon}\zeta_{\epsilon}+\int \int F^a_c Y^c_{[\alpha\alpha']}(\vec{p}\partial_a \vec{x}+h.c.)-2e_{[\alpha\alpha']\epsilon}z_{\epsilon}
\end{equation}
where
\begin{equation}  \label{26}
Y_{[\alpha\alpha']}^c=\frac{2}{\mu_{\alpha}\mu_{\alpha'}}(\nabla^b Y_{\alpha}\nabla_b \nabla^c Y_{\alpha'}-(\alpha \longleftrightarrow \alpha'))  ;
\end{equation}
while the second term in the last r.h.s. of (\ref{25}) is zero when
acting on the physical (constrained) Hilbert space, (\ref{25})
also
appears sandwiched between 2 momentum modes when trying to prove the
quantum analogue of (\ref{9}):
\[
[\zeta_{\alpha}p_{i\alpha}+p_{i\alpha}\zeta_{\alpha},\zeta_{\alpha'}p_{j\alpha'}+p_{j\alpha'}\zeta_{\alpha'}]
\]
\begin{equation}  \label{27}
=[\zeta_{\alpha}p_{i\alpha},\zeta_{\alpha'}p_{j\alpha'}]-[ \ , \ ]^{\dagger} + ([\zeta_{\alpha}p_{i\alpha},p_{j\alpha'}\zeta_{\alpha'}]-(i\leftrightarrow j)).
\end{equation}
Due to
\begin{equation}  \label{28}
[\zeta_{\alpha},\vec{p}_{\beta}]=i (d_{\alpha\beta\gamma}+e_{\alpha\beta\gamma})\vec{p}_{\gamma}
\end{equation}
and (25)
the first two terms of (\ref{27}) give
\[
\zeta_{\alpha}[p_{i\alpha},\zeta_{\alpha'}]p_{j\alpha'}+([\zeta_{\alpha},\zeta_{\alpha'}]p_{j\alpha'}+\zeta_{\alpha'}[\zeta_{\alpha},p_{j\alpha'}])p_{i\alpha}-h.c.
\]
\[
\approx -i(d_{\alpha'\alpha\epsilon}+e_{\alpha'\alpha\epsilon})\zeta_{\alpha}p_{i\epsilon}p_{j\alpha'}+i(e_{\alpha\alpha'\epsilon}-e_{\alpha'\alpha\epsilon})(\zeta_{\epsilon}-z_{\epsilon})p_{j\alpha'}p_{i\alpha}
\]
\begin{equation}  \label{29}
+ i(d_{\alpha \alpha' \epsilon}+ e_{\alpha\alpha'\epsilon})\zeta_{\alpha'}p_{j\epsilon}p_{i\alpha} - h.c.
\end{equation}
The two d-terms trivially cancel, and the $e\zeta pp$ terms due to $e_{\alpha(\beta\gamma)}=0$, resulting in only
\begin{equation}  \label{30}
-4i e_{[\alpha\alpha']\epsilon}z_{\epsilon}p_{i\alpha}p_{j\alpha'}
\end{equation}
remaining.

More difficult is the second part in (\ref{27}),
\begin{equation}  \label{31}
\zeta_{\alpha}p_{j\alpha'}[p_{i\alpha},\zeta_{\alpha'}]+([\zeta_{\alpha},p_{j\alpha'}]\zeta_{\alpha'}+p_{j\alpha'}[\zeta_{\alpha},\zeta_{\alpha'}])p_{i\alpha}-(i \leftrightarrow j).
\end{equation}
Writing in the first term $\zeta_{\alpha}p_{j\alpha'}=p_{j\alpha'}\zeta_{\alpha}+[\zeta_{\alpha},p_{j\alpha'}]$, and noting that
\begin{equation}  \label{32}
[\zeta_{\alpha},p_{j\alpha'}][p_{i\alpha},\zeta_{\alpha'}]-(i \leftrightarrow j)= [[\zeta_{\alpha},p_{j\alpha'}],[p_{i\alpha},\zeta_{\alpha'}]]=0
\end{equation}
(as each entry of the last commutator is a finite sum of momenta) all terms are of the form $p\zeta p$, namely:
\[
-ip_{j\alpha'}\zeta_{\alpha}(d_{\alpha'\alpha\epsilon}+
e_{\alpha'\alpha\epsilon})p_{i\epsilon}+i(d_{\alpha\alpha'\epsilon}+e_{\alpha\alpha'\epsilon})p_{j\epsilon}\zeta_{\alpha'}p_{i\alpha}
\]
\begin{equation}  \label{33}
+i p_{j\alpha'}(e_{\alpha\alpha'\epsilon}-e_{\alpha'\alpha\epsilon})(\zeta_{\epsilon}-z_{\epsilon})p_{i\alpha}+ip_{j\alpha'}\phi_{[\alpha\alpha']}p_{i\alpha}-(i \leftrightarrow j)
\end{equation}
where
\begin{equation}  \label{34}
\phi_{[\alpha\alpha']}=\int \int F^a_c(\varphi,\tilde{\varphi})Y^c_{[\alpha\alpha']}(\varphi)(\tilde\vec{p}\partial_a \vec{x}+ h.c.)(\tilde{\varphi}).
\end{equation}
While again the d-terms are canceling trivially, and the e-terms via
$e_{\alpha\beta\gamma}+e_{\alpha\gamma\beta}=0$, leaving again $-4i
e_{[\alpha\alpha']\epsilon}z_{\epsilon}p_{i\alpha}p_{j\alpha'}$, the potentially
most dangerous term is the one proportional to
\begin{equation}  \label{35}
p_{j\alpha'}\phi_{[\alpha\alpha']}p_{i\alpha}-(i \leftrightarrow j).
\end{equation}
While $[\phi,p]$ is linear in $p$ (hence possibly of the form (\ref{30})) it could be infinite. The following argument however renders (\ref{35}) finite (in fact: zero):
\[
2(p_{j\alpha'}\phi_{\alpha\alpha'}p_{i\alpha}-(i \leftrightarrow j) )
\]
\[
=\phi p_jp_i + [p_j,\phi]p_i +p_jp_i \phi + p_{j\alpha'}[\phi,p_i] + \phi p_ip_j+[p_{i\alpha},\phi_{\alpha\alpha'}]p_j
\]
\[
+p_{i\alpha}p_{j\alpha'}\phi_{\alpha\alpha'}+p_{i\alpha}[\phi_{\alpha\alpha'},p_{j\alpha'}]
\]
\begin{equation}  \label{36}
=2\phi_{\alpha\alpha'}p_{i\alpha}p_{j\alpha'}+2p_{i\alpha}p_{j\alpha'}\phi_{\alpha\alpha'}+[p_{i\alpha},[\phi_{\alpha\alpha'},p_{j\alpha'}]]+[p_{j\alpha'},[\phi_{\alpha\alpha'},p_{i\alpha}]]
\end{equation}
While the first two terms vanish between physical states, the inner
commutators in the last two terms are independent of $x$, causing (\ref{36}) to vanish for
each $\alpha$, $\alpha'$.

A similar analysis works for the terms
linear in $p$, and of degree $2M+1$ in $x$, as well as $\sim x_ix_j$ ($\rightarrow$ again indicating ``modulo terms containing the total momentum $\vec{P}$'' ; those terms have to work out just as in the classical theory, as $\vec{X}$ does not appear):
\[
[\int Y_{\alpha} \left( \frac{\vec{p}^2+g}{\rho} \right),\int Y_{\alpha'}(\tilde{\varphi})  \widetilde{\frac{\vec{p}^2+g}{\rho}}d^M\tilde{\varphi}]
\]
\[
=\int\int Y_{\alpha}(\varphi)Y_{\alpha'}(\tilde{\varphi})\left(\frac{\vec{p}}{\rho}[\vec{p},\tilde{g}]+[\vec{p},\tilde{g}]\frac{\vec{p}}{\rho}\right)-(\alpha \leftrightarrow \alpha')
\]
\[
=2i\int (Y_{\alpha}\partial_a Y_{\alpha'}-Y_{\alpha'}\partial_a Y_{\alpha})(\vec{p}\partial_b \vec{x}g g^{ab}+gg^{ab}\partial_b \vec{x}\cdot \vec{p})
\]
\[
=2i\int (Y_{\alpha}\partial_a Y_{\alpha'}-Y_{\alpha'}\partial_a Y_{\alpha})((\vec{p}\partial_b \vec{x}+\partial_b \vec{x} \vec{p})g g^{ab}+gg^{ab}(\partial_b \vec{x}\cdot \vec{p}+\vec{p}\partial_b \vec{x}))d^M \varphi
\]
\begin{equation}  \label{37}
\rightarrow i\int (Y_{\alpha}\partial_a Y_{\alpha'}-Y_{\alpha'}\partial_a Y_{\alpha})
((\partial_b \hat\zeta +  \phi_b)g g^{ab}+gg^{ab}(\partial_b \hat\zeta +  \phi_b))=[\mathcal{H}_{\alpha},\mathcal{H}_{\alpha'}]
\end{equation}
where
\begin{equation}  \label{38}
\phi_b:=\int F_b^c(\varphi,\tilde{\varphi})(\tilde\vec{p}\partial_c \vec{x}+\partial_c \vec{x} \cdot \tilde\vec{p})(\tilde{\varphi})d^M \tilde{\varphi},
\end{equation}
(\ref{24}) was used (with $z \equiv 0$), and the fact that $\vec{p}\partial_b \vec{x}$ and $\partial_b \vec{x} \cdot \vec{p}$ commute with
\begin{equation}  \label{39}
gg^{ab}=\frac{1}{(M-1)!}\epsilon^{aa_2 \ldots a_M}\epsilon^{bb_2 \ldots b_M}g_{a_2b_2}\ldots g_{a_Mb_M}
\end{equation}
- the reason being that
\begin{equation}  \label{40}
\vec{p}\partial_b \vec{x} g_{a_n b_n} = \sum_j p_j \partial_b x_j \partial_{a_n}x_j  \partial_{b_n}x_j + \sum_j\sum_{k\ne j} p_j \partial_b x_j \partial_{a_n}x_k \partial_{b_n}x_k;
\end{equation}
(the first sum vanishing upon contraction with $\epsilon^{bb_2 \ldots b_M}$, while in the second one all operators commute).

It is also easy to show that
\[
[p_{j\alpha'},\mathcal{H}_{\alpha}]=\int \int Y_{\alpha'}(\tilde{\varphi})Y_{\alpha(\varphi)}[p_j(\tilde{\varphi}),g(\varphi)]=2i \int Y_{\alpha'}\partial_a (Y_{\alpha} gg^{ab}\partial_b x_j)
\]
\begin{equation}  \label{41}
= -2i \int Y_{\alpha}\partial_a Y_{\alpha'}gg^{ab}\partial_b x_j
\end{equation}
Inserting (\ref{37}) and (\ref{41})
into
\[
x_{i\alpha}x_{j\alpha'}[\mathcal{H}_{\alpha},\mathcal{H}_{\alpha'}]+[\mathcal{H}_{\alpha},\mathcal{H}_{\alpha'}]x_{i\alpha}x_{j\alpha'}
\]
\[
+x_{i\alpha}[\mathcal{H}_{\alpha},\mathcal{H}_{\alpha'}]x_{j\alpha'}+x_{j\alpha'}[\mathcal{H}_{\alpha},\mathcal{H}_{\alpha'}]x_{i\alpha}
\]
\begin{equation}  \label{42}
x_{i\alpha}(\zeta_{\alpha'}[p_{j\alpha'},\mathcal{H}_{\alpha}]+[p_{j\alpha'},\mathcal{H}_{\alpha}]\zeta_{\alpha'})
+(\zeta_{\alpha'}[p_{j\alpha'},\mathcal{H}_{\alpha}]+[p_{j\alpha'},\mathcal{H}_{\alpha}]\zeta_{\alpha'})x_{i\alpha} - (i \leftrightarrow j)
\end{equation}
gives 16 terms of the form $x\partial x \partial \zeta g g^{ab}$
(in various orderings), from the $xx\mathcal{H}\mathcal{H}$ terms,
and $2\cdot 8$ $x\partial x \partial \zeta g g^{ab}$ terms (in
various orderings) from $[p,\mathcal{H}]x\zeta$ terms - which all
cancel each other (when using again that $\partial_a\zeta + \phi_b$
commutes with $gg^{ab}$, cp. (\ref{40})) - and the remaining terms
(involving the constraints) being (i times)
\begin{equation}  \label{43}
2x_i \partial x_j g g^{ab}\phi + 2\phi g g^{ab}x_i \partial x_j+  2x_i \phi \partial x_j g g^{ab} + 2 g g^{ab} \partial x_j \phi x_i - (i \leftrightarrow j).
\end{equation}
In analogy with (\ref{36}) the first two terms in (\ref{43}) vanish
between  physical states, while
\[
2 x_i \phi_b \partial_a x_j gg^{ab}+2gg^{ab}\partial_a x_j \phi_b x_i -  (i \leftrightarrow j)
\]
\[
=\phi_bx_i\partial_a x_j gg^{ab}+ [x_i,\phi_b]\partial_a x_j gg^{ab}+x_i [\phi_b,\partial_a x_j gg^{ab}]+x_i \partial_a x_j gg^{ab}\phi_b
\]
\[
+gg^{ab}\partial_a x_j [\phi_b,x_i]+ gg^{ab}\partial_a x_jx_i \phi_b + \phi_b gg^{ab}\partial_a x_jx_i + [gg^{ab}\partial_a x_j, \phi_b]x_i -(i \leftrightarrow j)
\]
\[
\approx [[x_i,\phi_b],\partial_a x_j gg^{ab}]+[x_i,[\phi_b,\partial_a x_j gg^{ab}]]- (i \leftrightarrow j)
\]
\[
=2[[x_i,\phi_b],\partial_a x_j gg^{ab}]- (i \leftrightarrow j) = 0
\]

Finally, in order to calculate all the terms containing $x_ip_j-(i
\leftrightarrow j)$ let us first note that (up to a possible
redefinition of $\mathbb{M}^2$ by a constant - which would, just
like in the standard treatment of strings, directly result in a
central extension) the classical relation (2), derived above, carries over identically to the
quantum case:
\begin{equation}  \label{44}
-i[\zeta_{\alpha},\mathcal{H}_{\beta}]=(3d_{\alpha\beta\gamma}+e_{\alpha\beta\gamma})\mathcal{H}_{\gamma}+4\delta_{\alpha\beta}(\mathbb{M}^2-c)+4\delta_{\alpha\beta}c.
\end{equation}
For the moment putting $c=0$, the remaining cross terms
\[
x_{i\alpha'}([\zeta_{\alpha},\mathcal{H}_{\alpha'}]p_{j\alpha}+p_{j\alpha}[\zeta_{\alpha},\mathcal{H}_{\alpha'}])
+
([\zeta_{\alpha},\mathcal{H}_{\alpha'}]p_{j\alpha}+p_{j\alpha}[\zeta_{\alpha},\mathcal{H}_{\alpha'}])x_{i\alpha'}
\]
\begin{equation}  \label{45}
+([\zeta_{\alpha},x_{i\alpha'}]p_{j\alpha}+p_{j\alpha}[\zeta_{\alpha},x_{i\alpha'}])\mathcal{H}_{\alpha'} +
\mathcal{H}_{\alpha'}([\zeta_{\alpha},x_{i\alpha'}]p_{j\alpha}+p_{j\alpha}[\zeta_{\alpha},x_{i\alpha'}])- (i \leftrightarrow j)
\end{equation}
together with $x[\mathcal{H},x]\mathcal{H}$ terms arising from $[\int (x_i\mathcal{H}+\mathcal{H}x_i),\int (x_j\mathcal{H}+\mathcal{H}x_j)]$, i.e.
\[
(x_{i\alpha}[\mathcal{H}_{\alpha},x_{j\alpha'}]\mathcal{H}_{\alpha'}+x_{j\alpha'}[x_{i\alpha},\mathcal{H}_{\alpha'}]\mathcal{H}_{\alpha})-(\ldots)^{\dagger}
\]
\begin{equation}  \label{46}
+([x_{i\alpha},\mathcal{H}_{\alpha'}]x_{j\alpha'}\mathcal{H}_{\alpha} + x_{i\alpha}\mathcal{H}_{\alpha'}[\mathcal{H}_{\alpha},x_{j\alpha'}] - (i \leftrightarrow j))
\end{equation}
give ($i$ times)
\[
[(3d_{\alpha\alpha'\epsilon}+e_{\alpha\alpha'\epsilon})\{x_{i\alpha'}(\mathcal{H}_{\epsilon}p_{j\alpha}+p_{j\alpha}\mathcal{H}_{\epsilon})+(\mathcal{H}_{\epsilon}p_{j\alpha}+p_{j\alpha}\mathcal{H}_{\epsilon})x_{i\alpha'}\}
\]
\[
+(-d_{\alpha\alpha'\epsilon}+e_{\alpha\alpha'\epsilon})\{(x_{i\epsilon}p_{j\alpha}+p_{j\alpha}x_{i\epsilon})\mathcal{H}_{\alpha'} +\mathcal{H}_{\alpha'}(x_{i\epsilon}p_{j\alpha}+p_{j\alpha}x_{i\epsilon})   \}
\]
\[
+4\delta_{\alpha\alpha'}\{x_{i\alpha'}(\mathbb{M}^2p_{j\alpha}+p_{j\alpha}\mathbb{M}^2) +(\mathbb{M}^2p_{j\alpha}+p_{j\alpha}\mathbb{M}^2)x_{i\alpha'} \})]  - (i \leftrightarrow j)
\]
\begin{equation}  \label{47}
-2d_{\alpha\alpha'\epsilon}\{x_{i\alpha'}p_{j\epsilon}\mathcal{H}_{\alpha}+\mathcal{H}_{\alpha}p_{j\epsilon}x_{i\alpha'} + p_{j\epsilon}x_{i\alpha'}\mathcal{H}_{\alpha} +
x_{i\alpha}\mathcal{H}_{\alpha'}p_{j\epsilon} -(i \leftrightarrow j) \}.
\end{equation}

While the e-terms trivially cancel (due to $e_{\alpha(\alpha'\epsilon)}=0$ and $e_{\alpha\alpha'\epsilon}[x_{i\alpha'},\mathcal{H}_{\epsilon}]=0$), and the $\mathbb{M}^2$ terms being (i times)
\begin{equation}  \label{48}
8(\mathbb{M}_{ij}\mathbb{M}^2+\mathbb{M}^2\mathbb{M}_{ij}),
\end{equation}
the d-terms can be seen to cancel by using
\[
d_{\alpha\alpha'\epsilon}[x_{i\alpha'},\mathcal{H}_{\epsilon}]p_{j\alpha} -(i \leftrightarrow j)
=2id_{\alpha\alpha'\epsilon}d_{\alpha'\epsilon\mu}p_{i\mu}p_{j\alpha}-(i \leftrightarrow j)=0
\]
(for each fixed $\alpha'$, $\epsilon$, due to $(\epsilon\mu)$ resp.
$[ij]$ (anti)symmetry) which allows to move $\mathcal{H}$ to the
very left or very right in the terms in (\ref{47}) where it appears
in the middle.

It then remains to check that
\[
-4i\eta^2
\left[X_iH+HX_i-2\zeta_0 P_i+2\frac{\mathbb{M}_{ik}P_{k}}{\eta},X_jH+HX_j-2\zeta_0 P_j+2\frac{\mathbb{M}_{jl}P_{l}}{\eta}\right]
\]
\begin{equation}  \label{49}
=-8(\mathbb{M}_{ij}\mathbb{M}^2+\mathbb{M}^2\mathbb{M}_{ij})
\end{equation}
and that the cross terms cancel
\begin{equation}  \label{50}
\left[X_iH+HX_i-2\zeta_0 P_i+2\frac{\mathbb{M}_{ik}P_{k}}{\eta},\int(x_j\mathcal{H}+\mathcal{H}x_j-\zeta p_j - p_j \zeta)\right] - (i \leftrightarrow j)
=0
\end{equation}

\section{Conclusion}
The above calculations provide a step towards the quantization of
higher-dimensional relativistic extended objects. While a
priori 
neither $M_{i-}$ , as defined above, nor
$\mathcal{H}_{\alpha}$, $\zeta_{\alpha}$, $\ldots$ need be
well-defined, the considerations presented in this paper (showing no
anomalies arising from the most naive ordering problems) should be
helpful when considering regularizations/renormalizations/product expansions of the
various operators involved.

\section*{Appendix}
The difficulty of resolving the ambiguities (and potential singularities) present in (commutators of) functionals of $\vec{x}$ and $\vec{p}$ is e.g. illustrated by the celebrated Virasoro algebra calculation; due to
\begin{equation}  \label{51}
[\alpha^i_m,\alpha^j_n]=m\delta_{m+n,0}
\end{equation}
being zero for $m+n \ne 0$ there are no ordering problems concerning
\begin{equation}  \label{52}
L_{n \ne 0} := \frac{1}{2}\sum_{k \in \mathbb{Z}} \vec{\alpha}_{n-k} \cdot \vec{\alpha}_k   ,
\end{equation}
while using that
\begin{equation}  \label{53}
[\vec{\alpha}_m,L_{n\ne 0}]=m \alpha_{n+m}
\end{equation}
one finds
\begin{equation}  \label{54}
[2L_{m\ne 0},L_{n\ne 0}]= \sum_{k \in \mathbb{Z}}\left(k \alpha_{m-k}\alpha_{k+n}+(m-k)\alpha_{m+n-k}\alpha_k\right) .
\end{equation}
Changing the summation index in the first sum, $k \to k-n$ would result in
\begin{equation}  \label{55}
(m-n)\sum_{k \in \mathbb{Z}}\vec{\alpha}_{m+n-k}\vec{\alpha}_k
\end{equation}
which is the desired answer $2(m-n)L_{m+n}$, if $m+n \ne 0$, but for
$n=-m$ would give
\[
2m \sum_{k \in \mathbb{Z}} \vec{\alpha}_{-k}\vec{\alpha}_k =
\]
\begin{equation}  \label{56}
2m (\sum_{k>0}\vec{\alpha}_{-k} \cdot
\vec{\alpha}_k+{\vec{\alpha}_0}^2+\sum_{k<0}(\vec{\alpha}_k \cdot
\vec{\alpha}_{-k}+[\vec{\alpha}_{-k},\vec{\alpha}_k]))  ,
\end{equation}
corresponding to the fact that
\begin{equation}  \label{57}
\left[\int_0^{2\pi}e^{i n \varphi}(\vec{p}^2+\vec{x}^{' \ 2}-\vec{p}\vec{x}^{'}-\vec{x}^{'}\vec{p}),\int_0^{2\pi}e^{-i n \tilde{\varphi}}(\vec{p}^2+\vec{x}^{' \ 2}-\vec{p}\vec{x}^{'}-\vec{x}^{'}\vec{p})(\tilde{\varphi})\right]
\end{equation}
by a naive calculation (missing $\delta^{'''}(\varphi,\tilde{\varphi})$) would be proportional to
\begin{equation}  \label{58}
\int_0^{2\pi} n(\vec{p}^2+\vec{x}^{' \ 2})d\varphi
\end{equation}
which does \emph{not} differ from ($n$ times)
\begin{equation}  \label{59}
2L_0 = \vec{\alpha}_0^2 + 2 \sum_{l>0}\vec{\alpha}_{-l}\vec{\alpha}_l
\end{equation}
by a finite constant proportional to $m(m^2-1)$ .

To actually obtain the desired $2mL_0 +\frac{2d}{12}m(m^2-1)$ one could proceed from (\ref{54})$_{n=-m}$ as follows:
\[
\sum_k \left(k \alpha_{m-k} \alpha_{k-m}+ (m-k)\alpha_{-k}\alpha_k \right)
\]
\[
=\sum_{k \ge m} k \alpha_{m-k}\alpha_{k-m} + \sum_{k<m}k (\alpha_{k-m}\alpha_{m-k}+[\vec{\alpha}_{m-k}, \vec{\alpha}_{k-m}])
\]
\[
+\sum_{k \ge 0} (m-k) \alpha_{-k}\alpha_{k} + \sum_{k<0}(m-k) (\alpha_{k}\alpha_{-k}+[\vec{\alpha}_{-k}, \vec{\alpha}_{k}])
\]
\[
=\sum_{l \ge 0}(m+l)\alpha_{-l} \alpha_l  + \sum_{l > 0}(m-l)\alpha_{-l} \alpha_l
+ \sum_{l \ge 0}(m-l)\alpha_{-l} \alpha_l  + \sum_{l > 0}(m+l)\alpha_{-l} \alpha_l
\]
\[
+ d \sum_{k=0}^{m-1}k(m-k)=
\]
\begin{equation}  \label{60}
= 2m\left(\vec{\alpha}_0^2+2 \sum_{l>0}\vec{\alpha}_{-l}\vec{\alpha}_l\right)+\frac{d}{6}m(m^2-1)
\end{equation}
i.e. changing summation indicies ($k=m+l$, $k=m-l$, $k=l$, $k=-l$ , respectively ) only in normal-ordered terms (and not in the commutator/c-number/ terms)

\vspace{4cm}

\section*{Acknowledgement}
I would like to thank G.Arutyunov, M.Bordemann, B.Durhuus, J.Goldstone, S.Theisen, and M.Trzetrzelewski for discussions.

\end{document}